\documentclass[a4paper]{llncs}
\usepackage{amssymb} 
\usepackage{amsmath}
\usepackage{marvosym}
\usepackage{algorithmic}
\usepackage{algorithm,float}
\usepackage{graphicx}
\usepackage{tikz}
\usepackage{graphicx}
\usepackage{subfig}
\usepackage{multirow}
\usepackage{hyperref}
\makeatletter
\newenvironment{breakablealgorithm}
{
	\begin{center}
		\refstepcounter{algorithm}
		\hrule height.1pt depth0pt \kern2pt
		\renewcommand{\caption}[2][\relax]{
			{\raggedright\textbf{\ALG@name~\thealgorithm} ##2\par}%
			\ifx\relax##1\relax 
			\addcontentsline{loa}{algorithm}{\protect\numberline{\thealgorithm}##2}%
			\else 
			\addcontentsline{loa}{algorithm}{\protect\numberline{\thealgorithm}##1}%
			\fi
			\kern2pt\hrule\kern2pt
		}
	}{
		\kern2pt\hrule\relax
	\end{center}
}
\makeatother
\begin{document}
\title{An Alternative Approach for Computing Discrete Logarithms in Compressed SIDH}
%
%
\author{Kaizhan Lin\inst{1} \and
	Weize Wang\inst{1} \and
	Lin Wang\inst{2} \and Chang-An Zhao\Letter\inst{1,3}}
%
%
\institute{School of Mathematics, Sun Yat-sen University,\\ Guangzhou 510275, P. R. China\\
	\email{linkzh5@mail2.sysu.edu.cn}\\
	\email{wangwz@mail2.sysu.edu.cn}\\
	\email{zhaochan3@mail.sysu.edu.cn}\\
	\and
	Science and Technology on Communication Security Laboratory,\\ Chengdu 610041, Sichuan, P. R. China\\
	\email{linwang@math.pku.edu.cn}
	\and
	Guangdong Key Laboratory of Information Security, \\ Guangzhou 510006, P. R. China
}

%
%

%
\maketitle

\begin{abstract}
	
	
	Currently, public-key compression of supersingular isogeny Diffie-Hellman (SIDH) and its variant, supersingular isogeny key encapsulation (SIKE) involve pairing computation and discrete logarithm computation. In this paper, we propose novel methods to compute only 3 discrete logarithms instead of 4, in exchange for computing a lookup table efficiently. The algorithms also allow us to make a trade-off between memory and efficiency. 
	Our implementation shows that the efficiency of our algorithms is close to that of the previous work, and our algorithms perform better in some special cases.

\end{abstract}

\section{Introduction} \label{sec1}
Isogeny-based cryptography has received widespread attention due to its small public key sizes in  post-quantum cryptography. The most attractive isogeny-based cryptosystems are supersingular isogeny Diffie-Hellman (SIDH)~\cite{SIDH} and its variant, supersingular isogeny key encapsulation (SIKE)~\cite{SIKE}. The latter one was submitted to NIST, and now it still remains one of the nine key encapsulation mechanisms in Round 3 of the NIST standardization process.

Indeed, Public key sizes in SIDH/SIKE can further be compressed. Azarderakhsh et al.~\cite{KeyCompression} firstly proposed a method for public-key compression, and later Costello et al.~\cite{EfficientCompression} proposed new techniques to further reduce the public-key size and make public-key compression practical. Zanon et al.~\cite{FasterIsogenyBased,Fasterkey} improved the implementation of compression and decompression by utilizing several techniques. Naehrig and Renes~\cite{dualisogeny} employed the dual isogeny to increase performance of compression techniques, while the methods for efficient binary torsion basis generation were presented in~\cite{xonly}. 

However, the implementation of pairing computation and discrete logarithm computation are still bottlenecks of public-key compression of SIDH/SIKE. Lin et al.~\cite{FasterPublicKeyCompression} saved about one third of memory for pairing computation and made it perform faster. To avoid pairing computation, Pereira and Barreto~\cite{WithoutPairings} compressed the public key with the help of ECDLP. As for discrete logarithms, Hutchinson et al.~\cite{MemoryOptimization} utilized signed-digit representation and torus-based representation to reduce the size of lookup tables for computing discrete logarithms. Both of them compress discrete logarithm tables by a factor of 2, and the former one reduces without any computational cost of lookup table construction. It makes practical to construct the lookup tables without precomputation.

In the current state-of-the-art implementation, there are $four$ values to be obtained in discrete logarithm computation. Note that one of the four values must be invertible in $\mathbb{Z}_{\ell^{e_\ell}}$. One only needs to get three new values~\cite{EfficientCompression}  by performing one inversion and three multiplications in $\mathbb{Z}_{\ell^{e_\ell}}$, and then transmit them. It is natural to ask whether one can compute the $three$ transmitted values directly during discrete logarithm computation.

In this paper, we propose an alternative way to compute discrete logarithms. We summarize our work as follows: 
\begin{itemize}
	\item  We propose a trick to compute only 3 discrete logarithms to compress the public key, in exchange for computing a lookup table efficiently.
	\item Currently, the algorithm used for discrete logarithm computation in compressed SIDH/SIKE is recursive. Inspired by~\cite{Parallelstrategies}, we present a non-recursive algorithm to compute discrete logarithms.
	\item We propose new algorithms to compute discrete logarithms in public-key compression of SIDH/SIKE. Our experimental results show that the efficiency of new algorithms is close to that of the previous work. Furthermore, our algorithms may perform well in storage constrained environments since we can make a memory-efficiency trade-off.
\end{itemize}

The sequel is organized as follows. In Section~\ref{sec2} we review the techniques that utilized for computing discrete logarithms in public-key compression. In Section~\ref{sec3} we propose new techniques to compute discrete logarithms without precomputation. We compare our experimental results with the previous work in Section~\ref{sec4} and conclude in Section~\ref{sec5}. 

\section{Notations and Preliminaries} \label{sec2}
\subsection{Notations}
In this paper, we use $E_A: y^2=x^3+Ax^2+x$ to denote a supersingular Montgomery curve defined over the field $\mathbb{F}_{p^2}=\mathbb{F}_{p}[i]/\langle i^2+1\rangle$, where $p=2^{e_2}3^{e_3}-1$. Let $E_6[2^{e_2}]=\langle P_2,Q_2\rangle$ and $E_6[3^{e_3}]=\langle P_3,Q_3\rangle$. We also use $\phi_2$ and $\phi_3$ to denote the $2^{e_2}$-isogeny and $3^{e_3}$-isogeny, respectively.  Besides, we define $\mu_n$ to be a multiplicative subgroup of order $n$ in $\mathbb{F}^*_{p^2}$, i.e,
\begin{equation*}\label{DLPgroup}
	\mu_{n}=\{\delta\in\mathbb{F}^*_{p^2}|\delta^{n}=1\}.
\end{equation*}

As usual, we denote the cost of one $\mathbb{F}_{p^2}$ field multiplication and squaring by $\textbf{M}$ and $\textbf{S}$, respectively. We also use $\textbf{m}$ and $\textbf{s}$ to denote the cost of one multiplication and squaring in the field $\mathbb{F}_p$. When estimating the cost, we assume that $\textbf{M}\approx 3\textbf{m}$, $S\approx 2\textbf{m}$ and $\textbf{s}\approx 0.8\textbf{m}$.
\subsection{Public-key Compression}\label{PKC}
In this subsection, we briefly review public-key compression of SIDH/SIKE, and concentrate on computing discrete logarithms.
We only consider how to compress two points of order $3^{e_3}$,
while the other case is similar. 
We refer to~\cite{SIDH,EfficientAlgorithms,Afastersoftware,SIKE} for more details of SIDH and SIKE. For their security analysis, see~\cite{OntheSecurity,ClawFinding,FasterAlgorithms,OntheCostof}.

Azarderakhsh et al.~\cite{KeyCompression} first presented techniques to compress the public key. The main idea is to generate a $3^{e_3}$-torsion basis $\langle U_3,V_3\rangle$ by a deterministic pseudo-random number generator, and then utilize this basis to linearly represent $\phi_2(P_3)$ and $\phi_2(Q_3)$. That is,
\begin{equation}\label{linearrepresentation}
	\begin{aligned}
			\left[\begin{array}{c}\phi_{2}\left(P_{3}\right) \\ \phi_{2}\left(Q_{3}\right)\end{array}\right]=\left[\begin{array}{ll}a_{0} & b_{0} \\ a_{1} & b_{1}\end{array}\right]\left[\begin{array}{l}U_3 \\ V_3\end{array}\right].
	\end{aligned}
\end{equation}
Note that 
\begin{equation}\label{fourpairing1}
	\begin{array}{l}\vspace{0.1cm}
		r_{0}=\mathrm{e}_{3^{e_3}}\left(U_3, V_3\right), \\\vspace{0.1cm}
		r_{1}=\mathrm{e}_{3^{e_3}}\left(U_3, \phi_{2}\left(P_{3}\right)\right)=\mathrm{e}_{3^{e_3}}\left(U_3, a_{0} U_3+b_{0} V_3\right)=r_{0}^{b_{0}}, \\\vspace{0.1cm}
		r_{2}=\mathrm{e}_{3^{e_3}}\left(U_3, \phi_{2}\left(Q_{3}\right)\right)=\mathrm{e}_{3^{e_3}}\left(U_3, a_{1} U_3+b_{1} V_3\right)=r_{0}^{b_{1}}, \\\vspace{0.1cm}
		r_{3}=\mathrm{e}_{3^{e_3}}\left(V_3, \phi_{2}\left(P_{3}\right)\right)=\mathrm{e}_{3^{e_3}}\left(V_3, a_{0} U_3+b_{0} V_3\right)=r_{0}^{-a_{0}}, \\\vspace{0.1cm}
		r_{4}=\mathrm{e}_{3^{e_3}}\left(V_3, \phi_{2}\left(Q_{3}\right)\right)=\mathrm{e}_{3^{e_3}}\left(V_3, a_{1} U_3+b_{1} V_3\right)=r_{0}^{-a_{1}}.
	\end{array}
\end{equation}
Therefore, with the help of bilinear pairings, one can compute $a_0$, $a_1$, $b_0$ and $b_1$ by computing four discrete logarithms in the multiplicative group $\mu_{3^{e_3}}$.

Instead of $\left(\phi_{2}\left(P_{3}\right), \phi_{2}\left(Q_{3}\right)\right)$, one could transmit the tuple $\left(a_0,b_0,a_1,b_1,A\right)$. Costello et al.~\cite{EfficientCompression} observed either $a_0\in \mathbb{Z}_{3^{e_3}}^*$ or $b_0\in \mathbb{Z}_{3^{e_3}}^*$ since the order of $\phi_{A}\left(P_{B}\right)$ is $3^{e_3}$, and concluded that the public key could be compressed to the tuple
\begin{equation*}
	\left(a_0^{-1}b_0,a_0^{-1}a_1,a_0^{-1}b_1,0,A\right) \mbox{, or } \left(b_0^{-1}a_0,b_0^{-1}a_1,b_0^{-1}b_1,1,A\right) \mbox{if $a_0\notin \mathbb{Z}^*_{3^{e_3}}$.}  
\end{equation*}

Zanon et al.~\cite{Fasterkey} proposed another new technique, called \textit{reverse basis decomposition}, to speed up the performance of computing discrete logarithms. Note that  $\langle\phi_{2}\left(P_{3}\right), \phi_{2}\left(Q_{3}\right)\rangle$ is also a $3^{e_3}$-torsion basis of $E_A$. The coefficient matrix in Equation~\eqref{linearrepresentation} is invertible, i.e.,

\begin{equation*}\label{reversebasis}
	\begin{aligned}
		\left[\begin{array}{c}U_3 \\ V_3\end{array}\right]=\left[\begin{array}{ll}c_{0} & d_{0} \\ c_{1} & d_{1}\end{array}\right]\left[\begin{array}{l}\phi_{2}\left(P_{3}\right) \\ \phi_{2}\left(Q_{3}\right)\end{array}\right]
	\end{aligned}, \text{ where }\begin{aligned} \left[\begin{array}{ll}c_{0} & d_{0} \\ c_{1} & d_{1}\end{array}\right]={\left[\begin{array}{ll}a_{0} & b_{0} \\ a_{1} & b_{1}\end{array}\right]}^{-1}\end{aligned}.
\end{equation*}

Correspondingly, the following pairing computation substitutes for Equation~\eqref{fourpairing1}:

\begin{equation}\label{DH2}
	\begin{aligned}\vspace{0.1cm}
		r_{0}&\!=\!\mathrm{e}_{3^{e_3}}\left(\phi_{2}\left(P_{3}\right), \phi_{2}\left(Q_{3}\right)\right)\!=\! \mathrm{e}_{3^{e_3}}\left(P_{3},Q_{3}\right)^{2^{e_2}}, \\\vspace{0.1cm}
		r_{1}&\!=\!\mathrm{e}_{3^{e_3}}\left(\phi_{2}\left(P_{3}\right), U_3\right)\!=\!\mathrm{e}_{3^{e_3}}\left(\phi_{2}\left(P_{3}\right), c_{0} \phi_{2}\left(P_{3}\right)+d_{0} \phi_{2}\left(Q_{3}\right)\right)\!=\!r_{0}^{d_{0}}, \\\vspace{0.1cm}
		r_{2}&\!=\!\mathrm{e}_{3^{e_3}}\left(\phi_{2}\left(P_{3}\right), V_3\right)\!=\!\mathrm{e}_{3^{e_3}}\left(\phi_{2}\left(P_{3}\right), c_{1} \phi_{2}\left(P_{3}\right)+d_{1} \phi_{2}\left(Q_{3}\right)\right)\!=\!r_{0}^{d_{1}}, \\\vspace{0.1cm}
		r_{3}&\!=\!\mathrm{e}_{3^{e_3}}\left(\phi_{2}\left(Q_{3}\right), U_3\right)\!=\!\mathrm{e}_{3^{e_3}}\left(\phi_{2}\left(Q_{3}\right), c_{0} \phi_{2}\left(P_{3}\right)+d_{0} \phi_{2}\left(Q_{3}\right)\right)\!=\!r_{0}^{-c_{0}}, \\\vspace{0.1cm}
		r_{4}&\!=\!\mathrm{e}_{3^{e_3}}\left(\phi_{2}\left(Q_{3}\right), V_3\right)\!=\!\mathrm{e}_{3^{e_3}}\left(\phi_{2}\left(Q_{3}\right), c_{1} \phi_{2}\left(P_{3}\right)+d_{1} \phi_{2}\left(Q_{3}\right)\right)\!=\!r_{0}^{-c_{1}}.
	\end{aligned}
\end{equation}
In this situation one needs to transmit
\begin{equation*}
	(-d_1^{-1}d_0,-d_1^{-1}c_1,d_1^{-1}c_0,0,A) \text{, or } (-d_0^{-1}d_1,d_0^{-1}c_1,-d_0^{-1}c_0,1, A) \text{ if $d_1\notin \mathbb{Z}^*_{3^{e_3}}$.}
\end{equation*}  

Since the value $r_0$ only depends on public parameters, the arbitrary order of $r_0$ could be precomputed to improve the implementation of computing discrete logarithms. In addition, note that the order of the group $\mu_{3^{e_3}}$ is smooth. Therefore, four discrete logarithms could be computed by using Pohlig-Hellman algorithm~\cite{PH} , as we will describe in the following subsection.

\subsection{Pohlig-Hellman algorithm}\label{PHalgorithm}
Pohlig-Hellman algorithm is an algorithm which is used to efficiently compute discrete logarithms in a group whose order is smooth. For a discrete logarithm $h=g^x\in\mu_{\ell^{e_\ell}}$, one could simplify it to $e_\ell$ discrete logarithms in a multiplicative group of order $\ell$.

\begin{algorithm}[H]
	\caption{Pohlig-Hellman Algorithm}
	\label{alg1}
	\begin{algorithmic}[1]
		\ENSURE $\langle g\rangle$: multiplicative group of order $\ell^{e_\ell}$; $h$: challenge.
		\REQUIRE $x$: integer $x\in [0,\ell^{e_\ell})$ such that $h=g^x$.
		\STATE $s\leftarrow g^{\ell^{e_\ell-1}}$, $x\leftarrow 0$, $h_0\leftarrow h$;
		\FOR{$i$ from $0$ to $e_\ell-1$}
		\STATE $t_i\leftarrow h_i^{\ell^{e_\ell-1-i}}$;
		\STATE \textbf{find} $x_i\in\{0,1,\cdots,\ell-1\}$ such that $t_i=s^{x_i}$;
		\STATE $x\leftarrow x+x_i\cdot\ell^i$, $h_{i+1} \leftarrow h_i\cdot g^{-x_i\ell^{i}}$;
		\ENDFOR
		\STATE \textbf{return} $x$.
	\end{algorithmic}
\end{algorithm}

As we can see in Algorithm~\ref{alg1}, a lookup table
\begin{equation*}
	T_1[i][j]=g^{-j\ell^{i}}, i = 0, 1, \cdots, e_\ell-1, j=0,1,\cdots,\ell-1, 
\end{equation*} 
can be precomputed to save the computational cost. Besides, one can also use a windowed version of Pohlig-Hellman algorithm to simplify the discrete logarithm to $\frac{e_{\ell}}{w}$ discrete logarithms in a group of order $L=\ell^w$, where $w|e_\ell$. The windowed version of Pohlig-Hellman algorithm reduces the loop length, but it consumes more storage.

When $w$ does not divide $e_\ell$ the procedure needs some modifications. Zanon et al. handled this situation by storing two tables~\cite[Section 6.2]{Fasterkey}:
\begin{equation}\label{twotables}
	\begin{array}{l}
		T_1[i][j]\!=\!g^{-j\ell^{wi}}, i = 0, 1, \cdots, e_\ell-1;\\
		\displaystyle T_2[i][j]=\left \{
		\begin{aligned}
			&\displaystyle g^{-j}, \mbox{if $i=0$},\\
			&\displaystyle g^{-j\ell^{w(i-1)+e_\ell\text{mod} w}}, \mbox{otherwise};
		\end{aligned}
	\right.
	\end{array}
\end{equation} 
where $j=0,1,\cdots,\ell^{e_\ell}-1$. This doubles the storage compared to the situation when $w$ divides $e_\ell$.

\subsection{Optimal Strategy}\label{optimalstrategy}
The time complexity of Algorithm~\ref{alg1} is $O(e_\ell^2)$. However, this strategy is far from optimal~\cite{Shoup}. Inspired by the optimal strategy of computing isogenies~\cite{SIDH}, Zanon et al.~\cite{Fasterkey} claimed that one can also adapt the optimal strategy into Pohlig-Hellman algorithm, reducing the time complexity to $O(e_\ell \log e_\ell)$ in the end.

Let $\triangle_n$ be a graph containing the vertices $\{\triangle_{j,k}|j+k\leq n-1,j\geq 0,k\geq0\}$, satisfying the following properties:
\begin{itemize}
	\item Each vertex $\triangle_{j,k}(j+k<n-1,j>0,k>0)$ has either two outgoing edges $\triangle_{j,k}\rightarrow \triangle_{j+1,k}$ and $\triangle_{j,k}\rightarrow \triangle_{j,k+1}$, or no edges at all; 
	\item Each vertex $\triangle_{j,0} (0<j<n-1)$ has only one outgoing edge $\triangle_{j,0}\rightarrow \triangle_{j+1,0}$, and $\triangle_{0,k} (0<k<n-1)$ has only one outgoing edge $\triangle_{0,k}\rightarrow \triangle_{0,k+1}$;
	\item Each vertex $\triangle_{j,k}(j+k=n-1)$ has no edges, called leaves; We also call the vertex $\triangle_{0,0}$ the root.
\end{itemize}

A subgraph is called a strategy if it contains a given vertex $\triangle_{j,k}$ such that all leaves and vertices can be reached
from $\triangle_{j,k}$. A strategy $\triangle^\prime_n$ of $\triangle_n$ is full if it contains the root $\triangle_{0,0}$ and all leaves $\triangle_{j,k}(j+k=n-1)$. Assigning the weights $p,q>0$ to the left edges and the right edges, respectively, \footnote{In this case, they are the cost of raising an element in $\mu_{p+1}$ to $\ell^w$-power and one multiplication in $\mathbb{F}_{p^2}$, respectively.} we can define the cost of an optimal strategy $\triangle_n^\prime$ by
\begin{equation}\label{costofstrategy}
	\hspace{-0.5cm}\displaystyle C_{p,q}(n)=\left \{
	\begin{aligned}
		&\displaystyle 0, \mbox{if $n=1$},\\
		&\displaystyle \min\left\{C_{p,q}(i)+C_{p,q}(n-i)+(n-i)p+iq\hspace{0.1cm}|\hspace{0.1cm} 0\leq i\leq n\right\}, \mbox{if $n>1$}.
	\end{aligned}
	\right.
\end{equation}

By utilizing Equation~\eqref{costofstrategy}, the optimal strategy could be attained by~\cite[Algorithm 6.2]{Fasterkey}.

\subsection{Signed-digit Representation}\label{signeddigit}
Hutchinson et al.~\cite{MemoryOptimization} reduced the memory for computing discrete logarithms by utilizing signed-digit representation~\cite{Signeddigits}. Here we only introduce the situation when $w$ divides $e_\ell$, while the other situation when $w$ does not divide $e_\ell$ one needs to store an additional table, but the handling is similar.

Instead of limiting $x=\log_g h\in\{0,1,\cdots,\ell^{e_\ell}-1\}$, we represent it by 
\begin{equation*}\label{signeddigitsequation}
	x=\sum_{k=0}^{e_\ell/w-1}D_k^\prime L^k,
\end{equation*}
where $L=\ell^w$ and $D_k^\prime\in [-\frac{\lceil L-1\rceil}{2},\frac{\lceil L-1\rceil}{2}]$. It seems that in this case we need to store
\begin{equation*}\label{singeddigittable1}
	T_1^{sgn}[i][j]=g^{jL^{i}},  i = 0, 1, \cdots, \frac{e_\ell}{w}-1, j\in[-\lceil\frac{ L-1}{2}\rceil,\lceil\frac{ L-1}{2}\rceil].
\end{equation*}

Since for any element $a+bi\in \mu_{p+1}$ $(a,b\in \mathbb{F}_p)$ and $p\equiv3\mod 4$,
\begin{equation*}\label{propertyofgroup}
	(a+bi)^{p+1}=1=(a+bi)(a+bi)^p=(a+bi)(a^p+b^pi^p)=(a+bi)(a-bi).
\end{equation*}

Hence, one inversion of an arbitrary element in $\mu_{p+1}$ is equal to its conjugate. This property guarantees one can reduce the table size by a factor of 2, i.e., 
\begin{equation*}\label{singeddigittable2}
	T_1^{sgn}[i][j]=g^{jL^{i}}, i = 0, 1, \cdots, \frac{e_\ell}{w}-1, j\in[1,\lceil\frac{ L-1}{2}\rceil].
\end{equation*}

\begin{remark}
	All the values in Column 0, i.e., $T_1^{sgn}[i][0]$, are equal to $g^0=1$. This is the reason why we do not need to precompute and store them.
\end{remark}

In fact Hutchinson et al. took advantages of torus-based representation of cyclotomic subgroup elements to further reduce the table size by a factor of 2. Since this technique is difficult to be utilized into this work, we do not review here and refer the interested reader to ~\cite{MemoryOptimization} for more details.
\subsection{Section Summary}\label{sectionsummary}
The implementation of computing discrete logarithms in  public-key compression of SIDH/SIKE has been optimized in recent years. However, it is still one of the main bottlenecks of key compression. 

To summarize, we propose Algorithm~\ref{alg2} to compute discrete logarithms by utilizing the techniques mentioned above.

\begin{algorithm}[h]
	\caption{Traverse($r$, $j$, $k$, $z$, $S$, $T_1^{sgn}$, $L$, $D$): 
		Improved Pohlig-Hellman Algorithm~\cite{Fasterkey}~\cite{MemoryOptimization}}
	\label{alg2}
	\begin{algorithmic}[1]
		\ENSURE $h$: value of root vertex $\triangle_{j,k}$ (i.e., challenge);
	    $j,k$: coordinates of root vertex $\triangle_{j,k}$;
	    $z$: number of leaves in subtree rooted at vertex $\triangle_{j,k}$;
		$S$: optimal strategy;
	    $T_1^{sgn}$: lookup table;
		$L$: $\ell^w$.
		\REQUIRE $D$: Array such that  $h=g^{\left(D[\frac{e_\ell}{w}-1]\cdots D[1]D[0]\right)_{L}}$.
		\IF{$z > 1$}
		\STATE $t\leftarrow S[z]$;
		\STATE $h^\prime\leftarrow h^{L^{z-t}}$;
		\STATE Traverse($h^\prime,j+(z-t),k,t,S,T_1^{sgn},L,D$);
		\STATE $h^\prime\leftarrow h\cdot \prod^{k+t-1}_{l=k}(T_1^{sgn}[j+l][|D[k]|-1])^{-sign(D[k])}$;
		\STATE Traverse($h^\prime,j,k+t,z-t,S,T_1^{sgn},L,D$);
		\ELSE
		\IF{$h=1$}
		\STATE $D[k]\leftarrow 0$.
		\ELSE
		\STATE \textbf{find} $x_k\in\{0,\cdots,\lfloor\frac{\ell^w-1}{2}\rfloor\}$ such that $h=T_1^{sgn}[\frac{e_\ell}{w}-1][x_k+1]$ or $h=\overline{T_1^{sgn}[\frac{e_\ell}{w}-1][x_k+1]}$;
		\IF{$h=T_1^{sgn}[\frac{e_\ell}{w}-1][x_k+1]$}
		\STATE $D[k]\leftarrow x_k+1$;
		\ELSE
		\STATE $D[k]\leftarrow -x_k-1$;
		\ENDIF
		\ENDIF
		\ENDIF
		\STATE \textbf{return} $D$.
	\end{algorithmic}
\end{algorithm}
%

\section{Computing Discrete Logarithms Without Precomputed Tables} \label{sec3}
As mentioned in Section~\ref{PKC}, one needs to compute four discrete logarithms in the multiplicative group $\langle r_0\rangle$ during public-key compression. Since $r_0$ is fixed, the techniques mentioned above are put to good use. 
In this section, we present another method to compute discrete logarithms, offering a time-memory trade-off as well.

\subsection{Three Discrete Logarithms}\label{TDL}
Note that the main purpose of computing discrete logarithms is to compute three values $(-d_1^{-1}d_0,-d_1^{-1}c_1,d_1^{-1}c_0)$ (or $(-d_0^{-1}d_1,d_0^{-1}c_1,-d_0^{-1}c_0)$ when $d_1$ is not invertible in $\mathbb{Z}_{\ell^{e_\ell}}$). For simplicity, we assume that $d_1$ is invertible and aim to compute $(-d_1^{-1}d_0,-d_1^{-1}c_1,d_1^{-1}c_0)$. 

Since $d_1$ is invertible in $\mathbb{Z}_{\ell^{e_\ell}}$, we can deduce that $r_2=r_0^{d_1}$ is a generator of the multiplicative group $\langle r_0\rangle$. Hence, instead of computing four discrete logarithms of $r_1,r_2,r_3,r_4$ to the base $r_0$ (defined in Equation~\eqref{DH2}), we consider three discrete logarithms of $r_1,r_3,r_4$ to the base $r_2$. It is clear that
\begin{equation*}\label{3DLP}
	\begin{array}{l}\vspace{0.1cm}
		\begin{aligned}\vspace{0.1cm}
			r_{1} &= r_{0}^{d_0}=r_{0}^{d_1\cdot d_1^{-1}\cdot d_0}=r_{2}^{d_1^{-1}d_0}, \\\vspace{0.1cm}
			r_{3} &= r_{0}^{c_0}=r_{0}^{d_1\cdot d_1^{-1}\cdot c_0}=r_{2}^{d_1^{-1}c_0}, \\\vspace{0.1cm}
			r_{4} &= r_{0}^{c_1}=r_{0}^{d_1\cdot d_1^{-1}\cdot c_1}=r_{2}^{d_1^{-1}c_1}. 
		\end{aligned}
	\end{array}
\end{equation*}

In other words, we only need to compute three discrete logarithms to compress the public key. Since it is unnecessary to compute $d_1^{-1}$ and multiply it by $d_0$, $c_0$ and $c_1$, we also save one inversion and three multiplications in $\mathbb{Z}_{\ell^{e_\ell}}$. Unfortunately, computing discrete logarithms to the base $r_0$ when lookup tables are available are much more efficient than computing discrete logarithms to the base $r_2$. Furthermore, it is impossible to precompute values to improve the performance due to the fact that the base $r_2$ depends on $d_1$. Hence, compared to the previous work in the case where $w|e_\ell$, one needs to efficiently construct the lookup table
\begin{equation}\label{singeddigittable3}
	T_1^{sgn}[i][j]=(r_2)^{(j+1)L^{i}}, i = 0, 1, \cdots, \frac{e_\ell}{w}-1, j=0,1,\cdots,\lceil\frac{ L-1}{2}\rceil-1.
\end{equation}

Zanon et al. handled the situation when $w\nmid e_\ell$ to precompute an extra lookup table, as described in Equation~\eqref{twotables}. Inspired by the method proposed by Pereira et al. when handling ECDLP~\cite[Section 4.4]{WithoutPairings}, we present a similar approach for computing discrete logarithms when $w\nmid e_\ell$. That is, instead of discrete logarithms of $r_1$, $r_3$, $r_4$ to the base $r_2$, we compute discrete logarithms of $(r_1)^{\ell^m}$, $(r_3)^{\ell^m}$, $(r_4)^{\ell^m}$ to the base $r_2$, where $m= e_\ell\hspace{-0.1cm}\mod w$. Correspondingly, the lookup table should be modified by the following:
\begin{equation*}\label{singeddigittable4}
	T_1^{sgn}[i][j]=r_2^{(j+1)L^{i}+\ell^m}, i = 0, 1, \cdots, \lfloor\frac{e_\ell}{w}\rfloor-1,  j=0,1,\cdots,\lceil\frac{ L-1}{2}\rceil-1.
\end{equation*}
In this situation, we recover the values $d_1^{-1}d_0(\text{ mod } \ell^{e_\ell-m})$, $d_1^{-1}c_0(\text{ mod } \ell^{e_\ell-m})$ and $d_1^{-1}c_1(\text{ mod } \ell^{e_\ell-m})$. Afterwards, we compute the three values as follows:
\begin{equation}\label{3additiveDLP}
	\begin{array}{l}\vspace{0.1cm}
		\begin{aligned}\vspace{0.1cm}
			r_1\cdot \left(r_2\right)^{-d_1^{-1}d_0\text{ mod } \ell^{e_\ell-m}}=\left( r_2\right) ^{d_1^{-1}d_0-\left(d_1^{-1}d_0\text{ mod } \ell^{e_\ell-m}\right)},\\\vspace{0.1cm}
			r_3\cdot \left( r_2\right) ^{-d_1^{-1}c_0\text{ mod } \ell^{e_\ell-m}}=\left( r_2\right) ^{d_1^{-1}c_0-\left(d_1^{-1}c_0\text{ mod } \ell^{e_\ell-m}\right)},\\\vspace{0.1cm}
			r_4\cdot \left( r_2\right) ^{-d_1^{-1}c_1\text{ mod } \ell^{e_\ell-m}}=\left( r_2\right) ^{d_1^{-1}c_1-\left(d_1^{-1}c_1\text{ mod } \ell^{e_\ell-m}\right)}.
		\end{aligned}
	\end{array}
\end{equation}
Finally, we compute three discrete logarithms of the above values to the base $(r_2)^{\ell^{e_\ell-m}}$ to recover the full digits of three values $-d_1^{-1}d_0$, $-d_1^{-1}c_1$ and $d_1^{-1}c_0$.

Since $\langle (r_2)^{\ell^{e_\ell-m}}\rangle$ is a multiplicative subgroup of $\langle (r_2)^{\ell^{e_\ell-w}}\rangle$, we can regard the last three discrete logarithms as the discrete logarithms to the base $(r_2)^{\ell^{e_\ell-w}}$, which are computed efficiently with the help of the lookup table. However, the computation in Equation~\eqref{3additiveDLP} is not an easy task. Therefore, except the construction of the lookup table, we also take into account how to obtain the three values mentioned in Equation~\eqref{3additiveDLP} with high efficiency when $w\nmid e_\ell$.


\subsection{Base Choosing}\label{BC}
Before constructing the lookup table, it is necessary to check whether $r_2$ is a generator of the multiplicative group $\langle r_0\rangle$. If not, we choose $r_1$ to be the base of discrete logarithms and construct the corresponding lookup table. 

Note that in this case, $d_1$ is unknown. So we can not determine the order of $r_2$ by computing the greatest common divisor of $d_1$ and $\ell^{e_\ell}$. Instead, we compute $(r_2)^{\ell^{e_\ell-1}}$ to check whether it is equal to 1. For any element $\delta=u+vi\in \mu_{p+1}$, we have
\begin{equation}\label{doutri}
	\begin{array}{l}\vspace{0.1cm}
		\begin{aligned}\vspace{0.1cm}
			\delta^2=&(u+vi)^2\\
			=&u^2-v^2+2uvi\\
			=&u^2-v^2+(1-u^2-v^2)i,\\
			\delta^3=&(u+vi)^3\\
			=&u^3+3u^2vi-3uv^2-v^3i\\
			=&u^3+3u^2\cdot vi-3u(1-u^2)-(1-u^2)\cdot vi\\
			=&-3u+4u^2\cdot u+(4u^2-1)\cdot vi.\\
		\end{aligned}
	\end{array}
\end{equation}
Hence, we can efficiently compute $(r_2)^{\ell^{e_\ell-1}}$ by squaring or cubing $e_\ell-1$ times with respect to $\ell$ and check whether it is equal to $1$.
Another advantage is that we also compute the values in the first column of the lookup table when $r_2$ is a generator of $\langle r_0\rangle$. Furthermore, when $r_2$ is a generator, the intermediate values
\begin{equation}\label{intermediatevaluesofr_2}
	C[i]=(r_2)^{\ell^i}, i=0,1,\cdots,e_\ell-m,
\end{equation}
could be utilized to speed up the performance when $w$ does not divide $e_\ell$. When $r_1$ is a generator, one can also construct the array 
\begin{equation*}\label{intermediatevaluesofr_1}
	C[i]=(r_1)^{\ell^i}, i=0,1,\cdots,e_\ell-m,
\end{equation*}
with a few additional square or cube operations. We will explain the reason why we also require these values in Section~\ref{DLC}.

We present Algorithm~\ref{alg3} for determining the base of discrete logarithms and computing the values in the first column of the lookup table. We also output the intermediate values that are used to improve the performance of discrete logarithms when $w\nmid e_\ell$.

\begin{breakablealgorithm}
	\caption{choose\_base($\ell$, $e_\ell$, $w$, $r_1$, $r_2$)}
	\label{alg3}
	\begin{algorithmic}[1]
		\ENSURE 
		$w:$ base power;
		$r_1,r_2$: elements defined in Equation~\eqref{DH2};
		$label$: sign bit used to mark the choice of the generator.\vspace{0.1cm}
		\REQUIRE $A$: values in the first column of the lookup table; $C$: intermediate values used to improve the performance of discrete logarithms when $w\nmid e_\ell$.
		\STATE  $label\leftarrow 1$, $A[0]\leftarrow r_2$, $C[0]\leftarrow r_2$, $j\leftarrow 0$;
		\FOR{$i$ from $0$ to $(e_\ell \mod w)-1$}
		\STATE $A[0]\leftarrow \left(A[0]\right)^\ell$, $j\leftarrow j+1$, $C[j]\leftarrow A[0]$;
		\ENDFOR
		\FOR{$i$ from $1$ to $\lfloor\frac{e_\ell}{w}\rfloor-1$}
		\STATE $A[i]\leftarrow A[i-1]$;
		\FOR{$k$ from $0$ to $w-1$}
		\STATE $A[i]\leftarrow \left(A[i]\right)^\ell$, $j\leftarrow j+1$, $C[j]\leftarrow A[i]$;
		\IF{$A[i]=1$}
		\STATE $label\leftarrow 0$, \textbf{break}.
		\ENDIF
		\ENDFOR
		\ENDFOR
		\IF{$label=1$}
		\STATE $t\leftarrow A[\lfloor\frac{e_\ell}{w}\rfloor-1]$;
		\FOR{$i$ from $0$ to $w-2$}
		\STATE $t\leftarrow t^\ell$, $j\leftarrow j+1$, $C[j]\leftarrow t$;
		\IF{$t=1$}
		\STATE $label\leftarrow 0$, \textbf{break}.
		\ENDIF
		\ENDFOR
		\ENDIF
		\IF{$label=0$}
		\STATE $A[0]\leftarrow r_1$, $C[0]\leftarrow r_1$, $j\leftarrow 0$;
		\FOR{$i$ from $0$ to $(e_\ell \mod w)-1$}
		\STATE $A[0]\leftarrow \left(A[0]\right)^\ell$, $j\leftarrow j+1$, $C[j]\leftarrow A[0]$;
		\ENDFOR
		\FOR{$i$ from $1$ to $\lfloor\frac{e_\ell}{w}\rfloor-1$}
		\STATE $A[i]\leftarrow A[i-1]$;
		\FOR{$k$ from $0$ to $w-1$}
		\STATE $A[i]\leftarrow \left(A[i]\right)^\ell$, $j\leftarrow j+1$, $C[j]\leftarrow A[i]$;
		\ENDFOR
		\ENDFOR
		\STATE $t\leftarrow A[\lfloor\frac{e_\ell}{w}\rfloor-1]$;
		\FOR{$i$ from $0$ to $w-m-1$}
		\STATE $t\leftarrow t^\ell$, $j\leftarrow j+1$, $C[j]\leftarrow t$;
		\ENDFOR
		\ENDIF
		\STATE \textbf{return} $label,A,C$.
	\end{algorithmic}
\end{breakablealgorithm}

\subsection{Lookup Table Construction}\label{LTC}
Algorithm~\ref{alg3} outputs the values in the first column of the lookup table. As we can see in Equation~\eqref{singeddigittable3}, all the values in the lookup table are the small powers of the values in the corresponding row. More precisely, 
\begin{equation*}\label{relationoflookuptable}
	T_1^{sgn}[i][j]=\left(T_1^{sgn}[i][1]\right)^{j+1},i = 0, 1, \cdots, \frac{e_\ell}{w}-1, j= 1,2,\cdots,\lceil\frac{ L-1}{2}\rceil-1.
\end{equation*}

Therefore, one can raise the powers of the values in the first column to generate all the values in the lookup table. As mentioned in Equation~\eqref{doutri}, the costs of squaring and cubing in the multiplicative group $\mu_{p+1}$ are approximately $2\textbf{s}\approx1.6\textbf{m}$ and $1\textbf{s}+2\textbf{m}\approx2.8\textbf{m}$, respectively. Both of them are more efficient than operating one multiplication in $\mathbb{F}_{p^2}$, which costs approximately 3\textbf{m}. Note that all the values are in the group $\mu_{p+1}$. One can utilize squaring and cubing operations, as we summarized in Algorithm~\ref{alg4}.

\begin{algorithm}
	\caption{T\_DLP($\ell$, $e_\ell$, $w$, $A$)}
	\label{alg4}
	\begin{algorithmic}[1]
		\ENSURE 
		$w:$ base power;
		$A$: values in the first column of the lookup table $T_1^{sgn}$.
		\REQUIRE $T_1^{sgn}$: entire lookup table.
		\FOR{$i$ from $0$ to $\lfloor\frac{e_\ell}{w}\rfloor-1$}
		\STATE $T_1^{sgn}[i][0]\leftarrow A[i]$;
		\ENDFOR
		\FOR{$i$ from $0$ to $\lfloor\frac{e_\ell}{w}\rfloor-1$}
		\FOR{$j$ from $1$ to $\lfloor\frac{\ell^w-1}{2}\rfloor$}
		\IF{$j\text{ mod } 2=1$}
		\STATE $T_1^{sgn}[i][j]\leftarrow \left(T_1^{sgn}[i][\frac{j-1}{2}]\right)^2$;
		\ELSE
		\IF{$j\text{ mod } 3=2$}
		\STATE $T_1^{sgn}[i][j]\leftarrow \left(T_1^{sgn}[i][\frac{j-2}{3}]\right)^3$;
		\ELSE
		\STATE $T_1^{sgn}[i][j]\leftarrow \left(T_1^{sgn}[i][\frac{j-1}{2}]\right)\cdot T_1^{sgn}[i][0]$;
		\ENDIF
		\ENDIF
		\ENDFOR
		\ENDFOR
		\STATE \textbf{return} $T_1^{sgn}$.
	\end{algorithmic}
\end{algorithm}

The bigger the base power $w$, the larger the size of the lookup table $T_1^{sgn}$, i.e., the higher the computational cost of lookup table construction, but the less discrete logarithms to be computed. Hence, just like efficiency-memory trade-offs provided by the previous work, we also explore the optimal base power $w$ to minimize the whole computational cost. We leave this exploration in Section~\ref{sec4}.

\subsection{Discrete Logarithm Computation}\label{DLC}
For ease of exposition, in this subsection we assume that we have chosen $r_2$ as the base of discrete logarithms. By utilizing Pohlig-Hellman algorithm, three discrete logarithms to the base $r_2$ could be simplified into discrete logarithms to the base $(r_2)^{\ell^{e_\ell-w}}$ or $(r_2)^{\ell^{e_\ell-m}}$. Indeed, the discrete logarithms to the base $(r_2)^{\ell^{e_\ell-m}}$ can also be regarded as discrete logarithms to the base $(r_2)^{\ell^{e_\ell-w}}$ since $(r_2)^{\ell^{e_\ell-m}}$ is an element in the multiplicative group $\langle(r_2)^{\ell^{e_\ell-w}}\rangle$. Thus, we consider how to compute discrete logarithms to the base $(r_2)^{\ell^{e_\ell-w}}$ first.

Note that all the entries in the last row of the lookup table $T_1^{sgn}$ are of the form
\begin{equation*}\label{lastrow}
	T_1^{sgn}[\lfloor\frac{e_\ell}{w}\rfloor-1][j]=(r_2)^{(j+1)\ell^{e_\ell-w}}, j=0,1, \cdots, \lceil\frac{L-1}{2}\rceil-1.
\end{equation*}

Thanks to signed-digit representation, all the entries in the last row of the lookup table and their conjugates consist of all nontrivial elements in the multiplicative group $\langle(r_2)^{\ell^{e_\ell-w}}\rangle$. Therefore, computing discrete logarithms to the base $(r_2)^{\ell^{e_\ell-w}}$ is relatively easy with the help of $T_1^{sgn}[\lfloor\frac{e_\ell}{w}\rfloor-1][j], j=0,1,\cdots,\lceil\frac{L-1}{2}\rceil-1$.

\begin{algorithm}[H]
	\caption{small\_DLP($\ell$, $w$, $h$, $B^\prime$)}
	\label{alg5}
	\begin{algorithmic}[1]
		\ENSURE 
		$w$: base power; $h$: challenge; $B^\prime$: last row of the lookup table $T_1^{sgn}$;
		\REQUIRE $x,sgn$: integers such that $h=\left(B^\prime[0]\right)^{sgn\cdot x}$. 
		\IF{$h=1$}
		\STATE $x\leftarrow 0$, $sgn\leftarrow 1$;
		\ELSE
		\STATE \textbf{find} $x\in\{0,\cdots,\lfloor\frac{L-1}{2}\rfloor\}$ such that $h=B^\prime[x]$ or $h=\overline{B^\prime[x]}$;
		\IF{$h=B^\prime[x]$}
		\STATE $x\leftarrow -x-1$;
		\ELSE
		\STATE $x\leftarrow x+1$;
		\ENDIF
		\ENDIF
		\STATE \textbf{return} $x$, $sgn$.
	\end{algorithmic}
\end{algorithm}
\begin{remark}\label{lastthreeDLPhandling}
	When handling discrete logarithms to the base $(r_2)^{\ell^{e_\ell-m}}$, the output of Algorithm~\ref{alg5} is $\ell^{w-m}$ times of the correct answer. Therefore, we should modify the output by dividing it by $\ell^{w-m}$.
\end{remark}

As we have pointed out in Section~\ref{TDL}, when the base power $w$ does not divide $e_\ell$, one efficiency issue to be solved is how to compute the values in Equation~\eqref{3additiveDLP}. We propose a method to deal with this issue by utilizing the intermediate values $C$ from Algorithm~\ref{alg3}.

In Algorithm~\ref{alg3}, we repeat squaring or cubing operations and store the intermediate values, as described in Equation~\eqref{intermediatevaluesofr_2}.
On the other hand, after computing discrete logarithms of $(r_1)^{\ell^m}$, $(r_3)^{\ell^m}$, $(r_4)^{\ell^m}$ to the base $r_2$, we recover $d_1^{-1}d_0(\text{ mod } \ell^{e_\ell-m})$, $d_1^{-1}c_0(\text{ mod } \ell^{e_\ell-m})$ and $d_1^{-1}c_1(\text{ mod } \ell^{e_\ell-m})$. Therefore, similar to the Double-and-Add algorithm, one can compute $r_2^{-d_1^{-1}d_0\text{ mod } \ell^{e_\ell-m}}$ (the other two are similar) according to the binary/ternary expansion of the value $-d_1^{-1}d_0(\text{ mod } \ell^{e_\ell-m})$, with respect to $\ell$. Afterwards, it just needs to perform three multiplications in $\mathbb{F}_{p^2}$ to obtain all the values.

In Algorithm~\ref{alg6}, we present pseudocode for computing $r_2^{-d_1^{-1}d_0\text{ mod } \ell^{e_\ell-m}}$, $r_2^{-d_1^{-1}c_0\text{ mod } \ell^{e_\ell-m}}$ and $r_2^{-d_1^{-1}c_1\text{ mod } \ell^{e_\ell-m}}$. Note that squaring in $\mu_{p+1}$ can also benefit from Equation~\eqref{doutri}. Hence, for Line 18 of Algorithm~\ref{alg6}, it would be efficient if we square $C[i]$ (or its conjugate $\overline{C[i]}$) first, and then perform a multiplication.

\begin{breakablealgorithm}
	\caption{fast\_power($\ell$, $w$, $D$, $C$)}
	\label{alg6}
	\begin{algorithmic}[1]
		\ENSURE $D$: array in base $L=\ell^w$ with signed digits; $C$: array from Algorithm~\ref{alg3};
		\REQUIRE $h$: $(C[0])^{\left(D[\lfloor\frac{ e_\ell}{w}\rfloor-2]\cdots D[1]D[0]\right)_L}$. 
		\STATE $h\leftarrow 1$, $i_1\leftarrow 0$, $i_2\leftarrow 0$;
		\FOR{$i$ from $0$ to $\lfloor\frac{ e_\ell}{w}\rfloor-2$}
		\STATE $t\leftarrow D[i]$, $s\leftarrow 1$;
		\IF{$D[i]<0$}
		\STATE $t\leftarrow -t$, $s\leftarrow -1$;
		\ENDIF
		\WHILE{$t>0$}
		\IF{$\ell=2$}
		\IF{$t\text{ mod }2=1$}
		\STATE $h\leftarrow h\cdot (C[i_2])^s$;
		\ENDIF
		\STATE $i_2\leftarrow i_2+1$, $t\leftarrow \lfloor \frac{t}{2}\rfloor$;
		\ELSE
		\IF{$t\text{ mod }3=1$}
		\STATE $h\leftarrow h\cdot (C[i_2])^s$;
		\ENDIF
		\IF{$t\text{ mod }3=2$}
		\STATE $h\leftarrow \left(C[i_2]\right)^{2s}\cdot h$;
		\ENDIF
		\STATE $i_2\leftarrow i_2+1$, $t\leftarrow \lfloor \frac{t}{3}\rfloor$;
		\ENDIF
		\ENDWHILE
		\STATE  $i_1\leftarrow i_1 + w$, $i_2\leftarrow i_1$;
		\ENDFOR
		\STATE \textbf{return} $h$.
	\end{algorithmic}
\end{breakablealgorithm}

It remains how to compute discrete logarithms of $r_1$, $r_3$ and $r_4$ to the base $r_2$ efficiently. Cervantes-V\'azquez et al. proposed a non-recursive algorithm to compute $\ell^{e_\ell}$-isogeny~\cite{Parallelstrategies}. Inspired by their work, we present Algorithm~\ref{alg7} to compute discrete logarithms.
Now we describe how Algorithm~\ref{alg7} works in detail. 
\vspace{0.3cm}

\textbf{Notations:} The input $h$ is the challenge of discrete logarithms, i.e, $r_1$, $r_3$ or $r_4$. The vector $Str$ is the linear representation of the optimal strategy. In the algorithm, we construct a stack, denoted by $Stack$, which contains the tuples of the form $(h_t, e_t, l_t)$, where $h_t\in \mu_{p+1}$ and $e_t,l_t\in\mathbb{N}$. Each tuple in $Stack$ represents the vertex which has been passed through (in left-first order), with the value $h_t$, the order $\ell^{e_\ell-e_t-m}$ and a right outgoing edge. When pushing a tuple into $Stack$, we also record the label $Str[i]$ of the previous vertex, denoted by $l_t$. The integers $(j,k)$ are coordinates of the last vertex which has been passed through. The other notations, such as the lookup table $T_1^{sgn}$, are defined as above.

\textbf{Lines 3-6:} As we described in Section~\ref{TDL}, we compute discrete logarithms of $(h)^{\ell^m}$ to the base $r_2$ when $w\nmid e_\ell$. So we first compute $(h)^{\ell^m}$ when $m\neq 0$. Afterwards, we push $((h)^{\ell^m},0,0)$ into $Stack$. 

\textbf{Lines 7-33:} This part is the core of Algorithm~\ref{alg7}. The main idea is to traverse the optimal strategy according to a left-first ordering and construct a stack to store all the vertices that have right outgoing edges. Once a discrete logarithm is computed, all the vertices in $Stack$ are replaced by their right vertices, respectively.

Line 7 checks if $k$ is equal to $\lfloor\frac{e_\ell}{w}\rfloor-1$, i.e, the rightmost vertex $\triangle_{0,\lfloor\frac{e_\ell}{w}\rfloor-1}$ has been traversed. In this case we jump out of the loop.

Line 8 aims to check whether the last vertex that has been passed through is a leaf or not. When the vertex is not a leaf, we go the left $Str[i]$ edges to enter the next split vertex and then \textbf{push} the information of this vertex into $Stack$ until the vertex is a leaf (Lines 10-13). When the vertex is a leaf, there are no edges to traverse left or right, and the values of the vertex is an element of order $\ell^w$ in the multiplicative group $\mu_{\ell^{e_\ell}}$. Hence, we \textbf{pop} the tuple from $Stack$ and then execute the algorithm \textbf{small\_DLP} in Lines 16-17. Then we store the result into the array $D$ in Lines 18-22.

Note that in this case, there are no left edges to be traversed. But all the right edges of the vertices in $Stack$ can be traversed since we have recovered $D[k]$. For each tuple $(h_t,e_t,l_t)$ in $Stack$, we execute $$h_t\leftarrow h_t\cdot \overline{T_1^{sgn}[e_t][x_t-1]} \textrm{~or~} h_t\leftarrow h_t\cdot T_1^{sgn}[e_t][x_t-1],$$ with respect to $sgn_t$ (Lines 23-31). 

The rest is to modify the position of the last vertex, as described in Line 32.

\textbf{Lines 34-40:} Now we have passed through the whole optimal strategy and in this case $Stack$ remains one tuple, i.e., it remains the vertex $\triangle_{0,\lfloor\frac{e_\ell}{w}\rfloor-1}$ that needed to be handled. Therefore, we \textbf{pop} the tuple from $Stack$ and execute the algorithm \textbf{small\_DLP} again. Finally, we store the answer into $D[k]$ (Note that $k=\lfloor\frac{e_\ell}{w}\rfloor-1$).

\textbf{Lines 41-50:} Line 41 checks whether the base power $w$ divides $e_\ell$. When $w$ divides $e_\ell$, we are done. If not, we need to compute the values in Equation~\eqref{3additiveDLP} and an extra discrete logarithm to the base $r_2^{\ell^{e_\ell-m}}$. Hence, when $m\neq 0$, we execute the algorithm \textbf{fast\_power} to compute $(r_2)^{\left(D[\lfloor\frac{ e_\ell}{w}\rfloor-2]\cdots D[1]D[0]\right)_L}$ with the help of the array $C$ and the efficiency of squaring and cubing in $\mu_{p+1}$. After that, we perform a multiplication in $\mathbb{F}_{p^2}$ and finally execute the algorithm \textbf{small\_DLP}. As we mentioned in Remark~\ref{lastthreeDLPhandling}, the output of \textbf{small\_DLP} is $\ell^{w-m}$ times of the correct answer. Therefore, we divide $\ell^{w-m}$ into the output.

\textbf{Line 51:} \textbf{Return} the array $D$.
\vspace{0.3cm}

Now we give a toy example to show how Algorithm~\ref{alg7} computes the discrete logarithm $h$ to the base $g$. For simplicity, we assume that $m=0$, and there are three leaves in the strategy $Str=(1,1)$, as illustrated in Figure (a). We first \textbf{push} the tuple $(h,0,0)$ into $Stack$. Now Lines 7-8 check that the vertex $\triangle_{0,0}$ is not a leaf, and therefore we are able to traverse left by squaring or cubing $w$ times and then \textbf{push} the tuple $(h^{\ell^w},1,1)$ into $Stack$, as described in Lines 10-13. Again, Line 8 checks that $\triangle_{0,1}$ is not a leaf as well, so we continue traversing left and \textbf{push} the tuple $(h^{\ell^w},2,1)$ into $Stack$ (Figure (c)). Note that $\triangle_{2,0}$ is a leaf of order $\ell^w$. We \textbf{pop} the tuple and then execute the algorithm \textbf{small\_DLP} to compute the discrete logarithm, and then we recover $D[0]$. Afterwards, Lines 23-31 handle all the vertices in $Stack$ by performing two multiplications in $\mathbb{F}_{p^2}$, as shown in Figures (d) and (e). In this case, we check that $\triangle_{1,1}$ is a leaf, so we \textbf{pop} the top tuple from $Stack$ and then execute \textbf{small\_DLP} again to recover $D[1]$. We traverse right from $\triangle_{0,1}$ to enter the rightmost vertex with the help of $D[1]$ (Figure (f)). Finally, Lines 34-40 \textbf{pop} the tuple and execute \textbf{small\_DLP} once again to recover $D[2]$.

\begin{figure*}[!htb]
	\centering
	\subfloat[]{
	\begin{tikzpicture}
		\draw  [green,line width=1.5](-0.5,-0.87)--(0,0);
		\draw  [green,line width=1.5](0.5,-0.87)--(0,0);
		\draw  [green,line width=1.5](1,-1.73)--(0.5,-0.87);
		\draw  [green,line width=1.5](-1,-1.73)--(-0.5,-0.87);
		\draw  [green,line width=1.5](0,-1.73)--(-0.5,-0.87);
		\filldraw [red] (0,0) circle (2pt);
		\filldraw [red] (-0.5,-0.87) circle (2pt);
		\filldraw [red] (0.5,-0.87) circle (2pt);
		\filldraw [red] (-1,-1.73) circle (2pt);
		\filldraw [red] (0,-1.73) circle (2pt);
		\filldraw [red] (1,-1.73) circle (2pt);
	\end{tikzpicture}
	}\hspace{0.5cm}
	\subfloat[]{
		\begin{tikzpicture}
			\draw  [blue,line width=1.5](-0.5,-0.87)--(0,0);
			\draw  [green,line width=1.5](0.5,-0.87)--(0,0);
			\draw  [green,line width=1.5](1,-1.73)--(0.5,-0.87);
			\draw  [green,line width=1.5](-1,-1.73)--(-0.5,-0.87);
			\draw  [green,line width=1.5](0,-1.73)--(-0.5,-0.87);
			\filldraw [red] (0,0) circle (2pt);
			\filldraw [red] (-0.5,-0.87) circle (2pt);
			\filldraw [red] (0.5,-0.87) circle (2pt);
			\filldraw [red] (-1,-1.73) circle (2pt);
			\filldraw [red] (0,-1.73) circle (2pt);
			\filldraw [red] (1,-1.73) circle (2pt);
		\end{tikzpicture}
	}\hspace{0.5cm}
	\subfloat[]{
		\begin{tikzpicture}
			\draw  [blue,line width=1.5](-0.5,-0.87)--(0,0);
			\draw  [green,line width=1.5](0.5,-0.87)--(0,0);
			\draw  [green,line width=1.5](1,-1.73)--(0.5,-0.87);
			\draw  [blue,line width=1.5](-1,-1.73)--(-0.5,-0.87);
			\draw  [green,line width=1.5](0,-1.73)--(-0.5,-0.87);
			\filldraw [red] (0,0) circle (2pt);
			\filldraw [red] (-0.5,-0.87) circle (2pt);
			\filldraw [red] (0.5,-0.87) circle (2pt);
			\filldraw [red] (-1,-1.73) circle (2pt);
			\filldraw [red] (0,-1.73) circle (2pt);
			\filldraw [red] (1,-1.73) circle (2pt);
		\end{tikzpicture}
	}\hspace{0.5cm}\\
	\subfloat[]{
		\begin{tikzpicture}
			\draw  [blue,line width=1.5](-0.5,-0.87)--(0,0);
			\draw  [blue,line width=1.5](0.5,-0.87)--(0,0);
			\draw  [green,line width=1.5](1,-1.73)--(0.5,-0.87);
			\draw  [blue,line width=1.5](-1,-1.73)--(-0.5,-0.87);
			\draw  [green,line width=1.5](0,-1.73)--(-0.5,-0.87);
			\filldraw [red] (0,0) circle (2pt);
			\filldraw [red] (-0.5,-0.87) circle (2pt);
			\filldraw [red] (0.5,-0.87) circle (2pt);
			\filldraw [red] (-1,-1.73) circle (2pt);
			\filldraw [red] (0,-1.73) circle (2pt);
			\filldraw [red] (1,-1.73) circle (2pt);
		\end{tikzpicture}
	}\hspace{0.5cm}
	\subfloat[]{
		\begin{tikzpicture}
			\draw  [blue,line width=1.5](-0.5,-0.87)--(0,0);
			\draw  [blue,line width=1.5](0.5,-0.87)--(0,0);
			\draw  [green,line width=1.5](1,-1.73)--(0.5,-0.87);
			\draw  [blue,line width=1.5](-1,-1.73)--(-0.5,-0.87);
			\draw  [blue,line width=1.5](0,-1.73)--(-0.5,-0.87);
			\filldraw [red] (0,0) circle (2pt);
			\filldraw [red] (-0.5,-0.87) circle (2pt);
			\filldraw [red] (0.5,-0.87) circle (2pt);
			\filldraw [red] (-1,-1.73) circle (2pt);
			\filldraw [red] (0,-1.73) circle (2pt);
			\filldraw [red] (1,-1.73) circle (2pt);
		\end{tikzpicture}
	}\hspace{0.5cm}
	\subfloat[]{
		\begin{tikzpicture}
			\draw  [blue,line width=1.5](-0.5,-0.87)--(0,0);
			\draw  [blue,line width=1.5](0.5,-0.87)--(0,0);
			\draw  [blue,line width=1.5](1,-1.73)--(0.5,-0.87);
			\draw  [blue,line width=1.5](-1,-1.73)--(-0.5,-0.87);
			\draw  [blue,line width=1.5](0,-1.73)--(-0.5,-0.87);
			\filldraw [red] (0,0) circle (2pt);
			\filldraw [red] (-0.5,-0.87) circle (2pt);
			\filldraw [red] (0.5,-0.87) circle (2pt);
			\filldraw [red] (-1,-1.73) circle (2pt);
			\filldraw [red] (0,-1.73) circle (2pt);
			\filldraw [red] (1,-1.73) circle (2pt);
		\end{tikzpicture}
	}
	\caption{A toy example of Algorithm~\ref{alg7}}
\end{figure*}
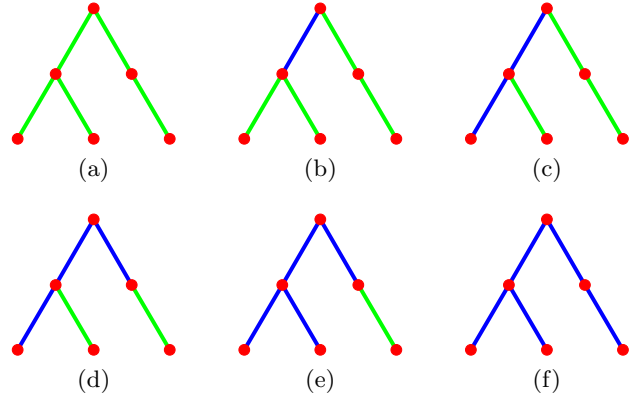

\vspace{0.5cm}
\begin{breakablealgorithm}
	\caption{PH\_DLP($\ell$, $e_\ell$, $w$, $h$, $Str$, $T_1^{sgn}$, $C$)}
	\label{alg7}
	\begin{algorithmic}[1]
		\ENSURE 
		$w$: base power; $h$: challenge; $Str$: Optimal strategy; $T_1^{sgn}$: entire lookup table, $C$: Array from Algorithm~\ref{alg3}; 
		\REQUIRE $D$: Array such that  $h=g^{\left(D[\lfloor\frac{e_\ell}{w}\rfloor-1]\cdots D[1]D[0]\right)_{\ell^w}}$.
		\STATE initialize a Stack $Stack$, which contains tuples of the form $(h_t, e_t, l_t)$, where $h_t\in\mu_{p+1}$, $e_t,l_t\in\mathbb{N}$.
		\STATE $B^\prime\leftarrow \text{last row of the lookup table}$ $T_1^{sgn}$, $i\leftarrow 0$, $j\leftarrow 0$, $k\leftarrow 0$, $m\leftarrow e_\ell\text{ mod }w$, $h_t\leftarrow h$;
		\FOR{$i_1$ from $0$ to $m-1$}
		\STATE $h_t\leftarrow \left( h_t\right) ^\ell$;
		\ENDFOR
		\STATE \textbf{Push} the tuple $(h_t,j,k)$ into $Stack$;
		\WHILE{$k\neq \lfloor \frac{e_\ell}{w}\rfloor-1$}
		\WHILE{$j+k\neq \lfloor \frac{e_\ell}{w}\rfloor-1$}
		\STATE $j\leftarrow j+Str[i]$;
		\FOR{$i_2$ from $0$ to $w\cdot Str[i]-1$}
		\STATE $h_t\leftarrow \left(h_t\right)^\ell$;
		\STATE \textbf{Push} the tuple $(h_t,j+k,Str[i])$ into $Stack$;
		\ENDFOR
		\STATE $i\leftarrow i+1$;
		\ENDWHILE
		\STATE \textbf{Pop} the top tuple $(h_t,e_t,l_t)$ from $Stack$;
		\STATE $(x_t, sgn_t)\leftarrow $ \textbf{small\_DLP}($\ell$, $w$, $h_t$, $B^\prime$);
		\IF{$sgn_t=1$}
		\STATE $D[k]\leftarrow x_t+1$;
		\ELSE
		\STATE $D[k]\leftarrow -x_t-1$;
		\ENDIF
		\FOR{each tuple $(h_t,e_t,l_t)$ in $Stack$}
		\IF{$x_t\neq 0$}
		\IF{$sgn_t=1$}
		\STATE $h_t\leftarrow h_t\cdot \overline{T_1^{sgn}[e_t][x_t-1]}$;
		\ENDIF
		\ELSE
		\STATE $h_t\leftarrow h_t\cdot T_1^{sgn}[e_t][x_t-1]$;
		\ENDIF
		\ENDFOR
		\STATE $j\leftarrow j-l_t$, $k\leftarrow k+1$;
		\ENDWHILE
		\STATE \textbf{Pop} the top tuple $(h_t,e_t,l_t)$ from $Stack$;
		\STATE $(x_t, sgn_t)\leftarrow $ \textbf{small\_DLP}($\ell$, $w$, $h_t$, $B^\prime$);
		\IF{$sgn_t=1$}
		\STATE $D[k]\leftarrow x_t+1$;
		\ELSE
		\STATE $D[k]\leftarrow -x_t-1$;
		\ENDIF
		\IF{$m\neq0$}
		\STATE $h_t\leftarrow$ \textbf{fast\_power}($\ell$, $D$, $C$);
		\STATE $h_t\leftarrow h\cdot\overline{h_t}$;
		\STATE $(x_t, sgn_t)\leftarrow$ \textbf{small\_DLP}($\ell$, $w$, $h_t$, $B^\prime$);
		\IF{$sgn_t=1$}
		\STATE $D[k+1]\leftarrow\frac{x_t+1}{\ell^{w-m}}$;
		\ELSE
		\STATE $D[k+1]\leftarrow -\frac{x_t+1}{\ell^{w-m}}$;
		\ENDIF
		\ENDIF
		\STATE \textbf{return} $D$.
	\end{algorithmic}
\end{breakablealgorithm}

\section{Cost Estimates and Implementation Results} \label{sec4}
In this section, we estimate the computational cost of discrete logarithms and compare our work with the previous work. We also report the implementation of key generation of SIDH by utilizing our techniques.
\subsection{Cost Estimates}
We neglect additions and mainly take into account multiplications and squarings (1\textbf{s} $\approx$ 0.8\textbf{m}) since they are much more expensive than additions. As shown in Table~\ref{tabce}, we predict that for all the Round-3 SIKE parameters, the cost of discrete logarithm computation in $\mu_{3^{e_3}}$ would be minimal when the base power $w$ is equal to 3. When handling $\mu_{2^{e_2}}$, the base power $w=4$ would be the best choice. 
\begin{table}[h]
	\begin{center}
		\caption{Cost estimates of three discrete logarithms utilizing our techniques. The minimal costs in the same row, i.e, in the same setting except the base power, are reported in bold.}\label{tabce}	
		\begin{tabular}{|c|c|c|c|c|c|c|} 
			\hline 
			\multicolumn{2}{|c|}{Setting} & w=1 & w=2 & w=3 & w=4 & w=6\\
			\hline  
			\multirow{2}*{SIKEp434}	& $\mu_{3^{e_3}}$ & 8892.6 & 6904.3 & \textbf{6463.3} & 7603 & 21915 \\
			& $\mu_{2^{e_2}}$ & 11762.4 & 7516 & 6083.6 & \textbf{5544.6} & 6232.4 \\
			\hline  
			\multirow{2}*{SIKEp503} & $\mu_{3^{e_3}}$ & 10780.3 & 8223.7 & \textbf{6859} & 8869.8 & 21960\\
			& $\mu_{2^{e_2}}$ & 13968.6 & 8902.2 & 8061.4 & \textbf{7441.7} & 8187.1 \\	
			\hline  
			\multirow{2}*{SIKEp610} & $\mu_{3^{e_3}}$ & 13477.5 & 9237.5 & \textbf{8552.2} & 9990.5 & 30941.8 \\
			& $\mu_{2^{e_2}}$ & 17650.2 & 12327.2 & 9542.4 & \textbf{9404.8} & 10256.6\\
			\hline  
			\multirow{2}*{SIKEp751} & $\mu_{3^{e_3}}$ & 17354.3 & 13265.9 & \textbf{12326.8} & 14076.5 & 39564.4\\
			& $\mu_{2^{e_2}}$ & 22181.4 & 14334.4 & 11594 & \textbf{10539} & 11552 \\
			\hline  
		\end{tabular} 
	\end{center}
\end{table}
\subsection{Implementation Results and Efficiency Comparisons}
Based on the Microsoft SIDH library\footnote[1]{ \url{https://github.com/Microsoft/PQCrypto-SIDH}} (version 3.4), we compiled our code
by using an 11th Gen Intel(R) Core(TM) i7-1185G7 @ 3.00GHz on 64-bit Linux.

For each setting we execute $10^4$ times and record the average cost of key generation, as summarized in Table~\ref{tabw}. The implementation results show that our prediction in the previous subsection is correct.
\begin{table}
	\begin{center}
		\caption{Implementation of key generation of compressed SIDH (expressed in millions of clock cycles). The minimal cost in the same row, i.e, in the same setting except the base power, are reported in bold. }\label{tabw}	
		\begin{tabular}{|c|c|c|c|c|c|c|} 
			\hline 
			\multicolumn{2}{|c|}{Setting} & \hspace{0.1cm}w=1\hspace{0.1cm} & \hspace{0.1cm}w=2\hspace{0.1cm} & \hspace{0.1cm}w=3\hspace{0.1cm} & \hspace{0.1cm}w=4\hspace{0.1cm} & \hspace{0.1cm}w=6\hspace{0.1cm}\\
			\hline  
			\multirow{2}*{SIKEp434} & $\mu_{3^{e_3}}$ & 6.41 & 6.16 & \textbf{6.07} & 6.27 & 8.68 \\
			& $\mu_{2^{e_2}}$ & 6.35 & 6.00 & 5.96 & \textbf{5.82} & 5.97 \\
			\hline  
			\multirow{2}*{SIKEp503} & $\mu_{3^{e_3}}$ & 8.50 & 8.27 & \textbf{7.93} & 8.54 & 11.93\\
			& $\mu_{2^{e_2}}$ & 8.51 & 8.25 & 8.10 & \textbf{8.01} & 8.12 \\
			\hline  
			\multirow{2}*{SIKEp610} & $\mu_{3^{e_3}}$ & 17.08 & 16.54 & \textbf{16.52} & 16.88 & 22.69 \\
			& $\mu_{2^{e_2}}$ & 16.31 & 15.89 & 15.65 & \textbf{15.59} & 15.71 \\
			\hline  
			\multirow{2}*{SIKEp751} & $\mu_{3^{e_3}}$ & 26.46 & 25.81 & \textbf{25.69} & 26.38 & 35.71 \\
			& $\mu_{2^{e_2}}$ & 27.26 & 26.36 & 25.84 & \textbf{25.32} & 26.09 \\
			\hline  
		\end{tabular} 
	\end{center}
\end{table}

On memory-constrained devices, our algorithms would be attractive for their relatively efficient performance even though we set small $w$. Table~\ref{tbmem} reports RAM requirements for the different parameters. 

\begin{table}
	\begin{center}
		\caption{RAM requirements (in KiB) for the different parameters.}\label{tbmem}	
		\begin{tabular}{|c|c|c|c|c|c|c|} 
			\hline 
			\multicolumn{2}{|c|}{Setting} & \hspace{0.1cm}w=1\hspace{0.1cm} & \hspace{0.1cm}w=2\hspace{0.1cm} & \hspace{0.1cm}w=3\hspace{0.1cm} & \hspace{0.1cm}w=4\hspace{0.1cm} & \hspace{0.1cm}w=6\hspace{0.1cm}\\
			\hline  
			\multirow{2}*{SIKEp434} & $\mu_{3^{e_3}}$ & 14.98 & 29.75 & 63.98 & 148.75 & 875.88 \\
			& $\mu_{2^{e_2}}$ & 23.63 & 23.63 & 31.50 & 47.25 & 126.00 \\
			\hline  
			\multirow{2}*{SIKEp503} & $\mu_{3^{e_3}}$ & 19.88 & 39.50 & 86.13 & 195.00 & 1183.00\\
			& $\mu_{2^{e_2}}$ & 31.25 & 31.25 & 41.50 & 62.00 & 164.00 \\
			\hline  
			\multirow{2}*{SIKEp610} & $\mu_{3^{e_3}}$ & 30.00 & 60.00 & 130.00 & 300.00 & 1820.00 \\
			& $\mu_{2^{e_2}}$ & 47.66 & 47.50 & 63.13 & 95.00 & 250.00 \\
			\hline  
			\multirow{2}*{SIKEp751} & $\mu_{3^{e_3}}$ & 44.81 & 89.25 & 192.56 & 442.50 & 2661.75 \\
			& $\mu_{2^{e_2}}$ & 69.75 & 69.75 & 93.00 & 139.50 & 372.00 \\
			\hline  
		\end{tabular} 
	\end{center}
\end{table}

Table~\ref{tabp} shows the comparison of efficiency between the previous work with ours. We can see that the efficiency of our algorithms is close to that of the previous work. When solving discrete logarithms in $\mu_{2^{e_2}}$, our algorithms are more efficient than the previous work when we set SIKEp434 or SIKEp751 as parameters. In addition, when the base power $w$ divides $e_\ell$, our algorithms perform better because there is no need to compute three values in Equation~\ref{3additiveDLP} and execute three additive discrete logarithms.

\begin{table}
	\begin{center}
		\caption{Key generation performance of the previous work and ours (expressed in millions of clock cycles). In the last column we report the ratio of the cost of the previous work to ours. In the same situation, we emphasize the lower cost in bold.}\label{tabp}	
		\begin{tabular}{|c|c|c|c|c|c|} 
			\hline 
			\multicolumn{2}{|c|}{Setting} & Previous work~\cite{SIKE} & This work & $w|e_\ell?$& Ratio\\
			\hline  
			\multirow{2}*{SIKEp434} & $\mu_{3^{e_3}}$ & \textbf{5.96} & 6.07 & No & 98.2\% \\
			& $\mu_{2^{e_2}}$ & 5.90  & \textbf{5.82} & Yes & 101.4\%\\
			\hline
			\multirow{2}*{SIKEp503} & $\mu_{3^{e_3}}$ & \textbf{8.07} & 8.14 & Yes & 99.14\% \\
			& $\mu_{2^{e_2}}$ & \textbf{7.93} & 8.01 & No & 99.00\% \\
			\hline
			\multirow{2}*{SIKEp610} & $\mu_{3^{e_3}}$ & \textbf{16.34} & 16.52 & Yes & 98.91\%\\
			& $\mu_{2^{e_2}}$ & \textbf{15.25} & 15.59 & No & 97.82\%\\
			\hline
			\multirow{2}*{SIKEp751} & $\mu_{3^{e_3}}$ & \textbf{25.20} & 25.69 & No & 98.09\%\\
			& $\mu_{2^{e_2}}$ & 25.61 & \textbf{25.32} & Yes & 101.15\% \\
			\hline
		\end{tabular} 
	\end{center}
\end{table}

\section{Conclusion} \label{sec5}
In this paper, we presented new techniques to compute discrete logarithms in public-key compression of SIDH/SIKE with no pre-computed tables. We analyze cost estimates of discrete logarithm computation with our techniques, and predict the best choices of $w$ in different situations. The implementation confirmed our deduction, and our algorithms to compute discrete logarithms in $\mu_{2^{e_2}}$ performed better in the situation when $w$ divides $e_2$. We believe that this work would be also attractive in storage restrained environments, for the reason that we can make a trade-off between memory and efficiency. 

Note that Algorithm~\ref{alg7} is a non-recursive algorithm. Hence, it would be more efficient in parallel environments. 
We leave those further explorations for future research.

\section*{Acknowledgments}
The authors thank the anonymous reviewers for their useful and valuable
comments. The work of Chang-An Zhao is partially supported by NSFC under Grant No. 61972428 and by the Major Program of Guangdong Basic and Applied Research under Grant No. 2019B030302008.
\bibliographystyle{splncs04}
\bibliography{ref}

\begin{thebibliography}{10}
\providecommand{\url}[1]{\texttt{#1}}
\providecommand{\urlprefix}{URL }
\providecommand{\doi}[1]{https://doi.org/#1}

\bibitem{OntheCostof}
Adj, G., Cervantes-V{\'a}zquez, D., Chi-Dom{\'i}nguez, J.J., Menezes, A.,
  Rodr{\'i}guez-Henr{\'i}quez, F.: On the {Cost} of {Computing} {Isogenies}
  {Between} {Supersingular} {Elliptic} {Curves}. In: Cid, C., Jacobson~Jr.,
  M.J. (eds.) Selected Areas in Cryptography -- SAC 2018. pp. 322--343.
  Springer International Publishing, Cham (2019)

\bibitem{Signeddigits}
Avizienis, A.: Signed-digit {Number} {Representations} for {Fast} {Parallel}
  {Arithmetic}. IRE Transactions on Electronic Computers  \textbf{EC-10}(3),
  389--400 (1961). \doi{10.1109/TEC.1961.5219227}

\bibitem{SIKE}
Azarderakhsh, R., Campagna, M., Costello, C., De~Feo, L., Hess, B., Hutchinson,
  A., Jalali, A., Jao, D., Karabina, K., Koziel, B., LaMacchia, B., Longa, P.,
  Naehrig, M., Pereira, G., Renes, J., Soukharev, V., Urbanik, D.:
  Supersingular {Isogeny} {Key} {Encapsulation} (2020), \url{http://sike.org}

\bibitem{KeyCompression}
{Azarderakhsh}, R., {Jao}, D., {Kalach}, K., {Koziel}, B., {Leonardi}, C.: Key
  {Compression} for {Isogeny-Based} {Cryptosystems}. In: Proceedings of the 3rd
  ACM International Workshop on ASIA Public-Key Cryptography. pp. 1--10 (2016)

\bibitem{Parallelstrategies}
Cervantes-Vázquez, D., Ochoa-Jiménez, E., Rodríguez-Henríquez, F.:
  {Parallel} {strategies} {for} {SIDH:} {Towards} {computing} {SIDH} {twice}
  {as} {fast}. Cryptology ePrint Archive, Report 2020/383 (2020),
  \url{https://ia.cr/2020/383}, accepted by IEEE Transactions on Computers

\bibitem{EfficientCompression}
Costello, C., Jao, D., Longa, P., Naehrig, M., Renes, J., Urbanik, D.:
  Efficient {Compression} of {SIDH} {Public} {Keys}. In: Coron, J.S., Nielsen,
  J.B. (eds.) Advances in Cryptology -- EUROCRYPT 2017. pp. 679--706. Springer
  International Publishing, Cham (2017)

\bibitem{EfficientAlgorithms}
Costello, C., Longa, P., Naehrig, M.: {Efficient} {Algorithms} for
  {Supersingular} {Isogeny} {Diffie-Hellman}. In: Robshaw, M., Katz, J. (eds.)
  Advances in Cryptology -- CRYPTO 2016. pp. 572--601. Springer Berlin
  Heidelberg, Berlin, Heidelberg (2016)

\bibitem{Afastersoftware}
Faz-Hernández, A., López, J., Ochoa-Jiménez, E., Rodríguez-Henríquez, F.:
  A {Faster} {Software} {Implementation} of the {Supersingular} {Isogeny}
  {Diffie-Hellman} {Key} {Exchange} {Protocol}. IEEE Transactions on Computers
  \textbf{67}(11),  1622--1636 (2018)

\bibitem{OntheSecurity}
Galbraith, S.D., Petit, C., Shani, B., Ti, Y.B.: On the {Security} of
  {Supersingular} {Isogeny} {Cryptosystems}. In: Cheon, J.H., Takagi, T. (eds.)
  Advances in Cryptology -- ASIACRYPT 2016. pp. 63--91. Springer Berlin
  Heidelberg, Berlin, Heidelberg (2016)

\bibitem{MemoryOptimization}
Hutchinson, A., Karabina, K., Pereira, G.: {Memory} {Optimization} {Techniques}
  for {Computing} {Discrete} {Logarithms} in {Compressed} {SIKE}. In: Cheon,
  J.H., Tillich, J.P. (eds.) Post-Quantum Cryptography. pp. 296--315. Springer
  International Publishing, Cham (2021)

\bibitem{SIDH}
Jao, D., De~Feo, L.: Towards {Quantum-Resistant} {Cryptosystems} from
  {Supersingular} {Elliptic} {Curve} {Isogenies}. In: Yang, B.Y. (ed.)
  Post-Quantum Cryptography. pp. 19--34. Springer Berlin Heidelberg, Berlin,
  Heidelberg (2011)

\bibitem{ClawFinding}
Jaques, S., Schanck, J.M.: Quantum {Cryptanalysis} in the {RAM} {Model}:
  {Claw-Finding} {Attacks} on {SIKE}. In: Boldyreva, A., Micciancio, D. (eds.)
  Advances in Cryptology -- CRYPTO 2019. pp. 32--61. Springer International
  Publishing, Cham (2019)

\bibitem{FasterPublicKeyCompression}
Lin, K., Lin, J., Wang, W., an~Zhao, C.: Faster {Public-key} {Compression} of
  {SIDH} with {Less} {Memory}. Cryptology ePrint Archive, Report 2021/992
  (2021), \url{https://ia.cr/2021/992}

\bibitem{dualisogeny}
Naehrig, M., Renes, J.: Dual {Isogenies} and {Their} {Application} to
  {Public-Key} {Compression} for {Isogeny-Based} {Cryptography}. In: Galbraith,
  S.D., Moriai, S. (eds.) Advances in Cryptology -- ASIACRYPT 2019. pp.
  243--272. Springer International Publishing, Cham (2019)

\bibitem{WithoutPairings}
Pereira, G.C.C.F., Barreto, P.S.L.M.: {Isogeny-Based} {Key} {Compression}
  {Without} {Pairings}. In: Garay, J.A. (ed.) Public-Key Cryptography -- PKC
  2021. pp. 131--154. Springer International Publishing, Cham (2021)

\bibitem{xonly}
Pereira, G.C.C.F., Doliskani, J., Jao, D.: $x$-only point addition formula and
  faster compressed {SIKE}. Journal of Cryptographic Engineering  \textbf{11},
  57--69 (2021)

\bibitem{FasterAlgorithms}
Petit, C.: Faster {Algorithms} for {Isogeny} {Problems} {Using} {Torsion}
  {Point} {Images}. In: Takagi, T., Peyrin, T. (eds.) Advances in Cryptology --
  ASIACRYPT 2017. pp. 330--353. Springer International Publishing, Cham (2017)

\bibitem{PH}
Pohlig, S., Hellman, M.: {An} {Improved} {Algorithm} for {Computing}
  {Logarithms} over {GF($p$)} and {Its} {Cryptographic} {Significance}
  {(Corresp.)}. IEEE Trans. Inf. Theor.  \textbf{24}(1),  106–110 (2006)

\bibitem{Shoup}
Shoup, V.: A computational introduction to number theory and algebra. Cambridge
  University Press (2005)

\bibitem{FasterIsogenyBased}
Zanon, G.H.M., Simplicio, M.A., Pereira, G.C.C.F., Doliskani, J., Barreto,
  P.S.L.M.: {Faster} {Isogeny-Based} {Compressed} {Key} {Agreement}. In: Lange,
  T., Steinwandt, R. (eds.) Post-Quantum Cryptography. pp. 248--268. Springer
  International Publishing, Cham (2018)

\bibitem{Fasterkey}
Zanon, G.H.M., Simplicio, M.A., Pereira, G.C.C.F., Doliskani, J., Barreto,
  P.S.L.M.: {Faster} {Key} {Compression} for {Isogeny-Based} {Cryptosystems}.
  IEEE Transactions on Computers  \textbf{68}(5),  688--701 (2019)

\end{thebibliography}
\end{document}